\documentstyle[twocolumn,aps,epsfig,here]{revtex}

\begin{document}

\title{Energy dependence of particle multiplicities in central Au+Au collisions} 
\author { B.B.Back$^1$, M.D.Baker$^2$, 
D.S.Barton$^2$, R.R.Betts$^6$, R.Bindel$^7$,  
A.Budzanowski$^3$, W.Busza$^4$, A.Carroll$^2$,
J.Corbo$^2$, M.P.Decowski$^4$, 
E.Garcia$^6$, N.George$^1$, K.Gulbrandsen$^4$, 
S.Gushue$^2$, C.Halliwell$^6$, 
J.Hamblen$^8$, C.Henderson$^4$, D.Hicks$^2$, 
D.Hofman$^6$, R.S.Hollis$^6$, R.Ho\l y\'{n}ski$^3$, 
B.Holzman$^2$, A.Iordanova$^6$,
E.Johnson$^8$, J.Kane$^4$, J.Katzy$^{4,6}$, 
N.Khan$^8$, W.Kucewicz$^6$, P.Kulinich$^4$, C.M.Kuo$^5$,
W.T.Lin$^5$, S.Manly$^{8}$,  D.McLeod$^6$, J.Micha\l owski$^3$,
A.Mignerey$^7$, J.M\"ulmenst\"adt$^4$, R.Nouicer$^6$, 
A.Olszewski$^{3}$, R.Pak$^2$, I.C.Park$^8$, 
H.Pernegger$^4$, M.Rafelski$^2$, M.Rbeiz$^4$, C.Reed$^4$, L.P.Remsberg$^2$, 
M.Reuter$^6$, C.Roland$^4$, G.Roland$^4$, L.Rosenberg$^4$, 
J. Sagerer$^6$, P.Sarin$^4$, P.Sawicki$^3$, 
W.Skulski$^8$, 
S.G.Steadman$^4$, P.Steinberg$^2$,
G.S.F.Stephans$^4$,  M.Stodulski$^3$, A.Sukhanov$^2$, 
J.-L.Tang$^5$, R.Teng$^8$, A.Trzupek$^3$, 
C.Vale$^4$, G.J.van Nieuwenhuizen$^4$, 
R.Verdier$^4$, B.Wadsworth$^4$, F.L.H.Wolfs$^8$, B.Wosiek$^3$, 
K.Wo\'{z}niak$^{2,3}$, 
A.H.Wuosmaa$^1$, B.Wys\l ouch$^4$\\
(PHOBOS Collaboration) \\
$^1$ Physics Division, Argonne National Laboratory, Argonne, IL 60439-4843\\
$^2$ Chemistry and C-A Departments, Brookhaven National Laboratory, Upton, NY 11973-5000\\
$^3$ Institute of Nuclear Physics, Krak\'{o}w, Poland\\
$^4$ Laboratory for Nuclear Science, Massachusetts Institute of Technology, Cambridge, MA 02139-4307\\
$^5$ Department of Physics, National Central University, Chung-Li, Taiwan\\
$^6$ Department of Physics, University of Illinois at Chicago, Chicago, IL 60607-7059\\
$^7$ Department of Chemistry, University of Maryland, College Park, MD 20742\\
$^8$ Department of Physics and Astronomy, University of Rochester, Rochester, NY 14627\\
}
\date{\today}
\maketitle

\begin{abstract}\noindent
We present the first measurement of the pseudorapidity density  of 
primary charged particles in Au+Au collisions 
at $\sqrt{s_{_{NN}}} =$ 200~GeV. For the 6\% most central collisions,
we obtain $dN_{ch}/d\eta |_{|\eta|<1} = 650 \pm 35 \mbox{(syst)}$. 
Compared to collisions at $\sqrt{s_{_{NN}}} =$ 130~GeV, the highest 
energy studied previously,  an increase by a factor of $1.14 \pm 0.05$ 
is found.
The energy dependence of the pseudorapidity density
is discussed in comparison with data from proton-induced collisions 
and theoretical predictions. 
\end{abstract}


Collisions of gold nuclei at an energy of $\sqrt{s_{_{NN}}} =$ 200~GeV
have been studied using the PHOBOS detector. 
PHOBOS is one of the experiments at the Relativistic Heavy-Ion Collider 
(RHIC) at Brookhaven National Laboratory 
aimed at understanding the behavior of strongly interacting 
matter at high temperature and density. Quantum chromodynamics (QCD), 
the fundamental theory of strong interactions, predicts that under 
these conditions, which may be probed in heavy-ion collisions, a new 
state of matter will be formed, the quark-gluon 
plasma (QGP)\cite{qgp}. In this state, quarks and gluons are no longer 
confined inside hadrons, as is the case for normal nuclear matter.
Information about the particle and energy density achieved in the early stages
of the collision process is carried by
the pseudorapidity  density of particles emitted from the primary collision point \cite{bjorken}.
In this analysis, we have determined the pseudorapidity density  of 
charged particles, $dN_{ch}/d\eta$, in the most central Au+Au collisions. 
We focused in particular on the region near 
$\eta=0$, where $\eta = - \ln \tan(\theta/2)$ and  
$\theta$ is the polar angle from the beam axis. 

In combination with results from lower energies, these data permit a 
systematic analysis of particle production mechanisms in nucleus-nucleus 
collisions. Extension of the energy range to $\sqrt{s_{_{NN}}} = $ 200~GeV
allows a study of the relative contributions of hard parton-parton 
scattering processes, which can be calculated using perturbative QCD,
and soft processes, which are treated by phenomenological models
that describe the non-perturbative sector of QCD. With increasing 
collision energy, hard processes 
are expected to become increasingly important for particle production
near mid-rapidity.

For Au+Au collisions at RHIC energies,
the yield and momentum distribution of 
particles produced by
hard scattering processes may be modified by ``jet quenching'', 
{\it i.e.}\ the energy loss of high momentum partons in the nuclear medium
\cite{jet_quench_theory}. 
This phenomenon has been proposed as a diagnostic tool for characterizing the 
initial parton density in Au+Au collisions at these energies. Preliminary 
results indicate that in central Au+Au collisions at $\sqrt{s_{_{NN}}} = $ 130~GeV the 
particle spectra at large transverse 
momenta, normalized to $p\overline{p}$ collisions, indeed change in comparison with
lower collision energies \cite{jet_quench_qm}.

Early predictions for the charged particle pseudorapidity density varied 
by more than a factor of two, as shown in \cite{armesto_0002163}.
Data on the primary charged particle density
$dN_{ch}/d\eta |_{|\eta|<1}$ at energies of $\sqrt{s_{_{NN}}} = $ 56 and 
130~GeV \cite{phobosprl} have been analyzed in a wide variety of 
theoretical models \cite{wang_0008014,barshay_0104303,accardi_0104060,jeon_0009032,eskola_0106330,dias_de_deus_0008086,kahana_0010043,armesto_0104269}. 
Generally, most models allowed a reasonable description of the energy 
dependence of $dN_{ch}/d\eta |_{|\eta|<1}$ up to $\sqrt{s_{_{NN}}} =$ 130~GeV,
with suitable choices of parameters. 
For the ratio of particle densities near $\eta=0$ at 130~GeV and 200~GeV, most
calculations predict an increase between 9 and 20\%. 
An interesting exception is the model of Wang and Gyulassy 
\cite{wang_0008014}. When the effects of jet quenching are included in 
their calculation, they not only predict a suppression of 
particle spectra at high transverse momenta, but also a
change of the energy dependence of $dN_{ch}/d\eta |_{|\eta|<1}$,
leading to an increase of more than 30\% between 130 and 200 GeV when
the default parameter set is used.
The energy dependence of particle production presented here provides 
important constraints on this effect.

Details of the PHOBOS experimental setup can be found 
elsewhere\cite{phobos1,phobos2}. The apparatus, shown schematically in Fig.~1, employs 
silicon detectors for 
vertex finding, particle tracking and multiplicity measurements. This 
analysis is based on the first 6 layers of the 16 layer two-arm 
spectrometer (SPEC), the two-layer vertex detector (VTX), the single-layer 
octagon barrel detector (OCT) and the three single-layer ring detectors (RING)
located on either side of the interaction point. 
The acceptance of SPEC, VTX and OCT 
includes $-1 < \eta < 1$, covering different regions of azimuth.
The combined acceptance of OCT and RING detectors reaches $\eta = \pm 5.4$.
For the 2001 run period, 137168 silicon channels were read out, of 
which less than 2\% were non-functional. 

The detector setup also included two sets of 16 scintillator 
counters (``paddle counters'') located at $-3.21$~m (PN) and 
$3.21$~m (PP) relative to the nominal interaction  point along the 
beam ($z$) axis. These counters covered pseudorapidities between
$3 < |\eta |< 4.5$ and served as the primary event trigger. In combination 
with  the zero-degree calorimeters at $z = \pm 18.5$~m,
which measured the energy deposited by spectator neutrons, PP and PN were also
used for offline event selection.

Monte Carlo (MC) simulations of the detector performance 
were based on the HIJING event generator \cite{hijing} 
and the GEANT~3.21 simulation package, folding in
the signal response for scintillator counters and silicon sensors.
In the calibration of the analysis methods we employed a particle 
selection procedure that allowed us to iteratively modify the MC
output and optimize the agreement of the deposited energy 
distributions in data and MC.

\begin{figure}[t]
\centerline{
\epsfig{file=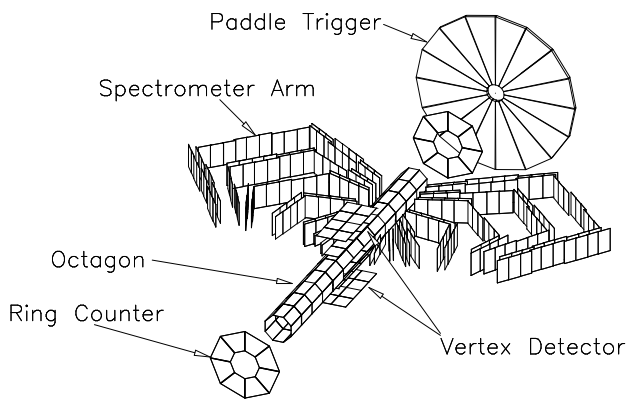,width=8.5cm}}
\caption{Detector setup for the 2001 running period. 
Shown are the active elements of the detector systems used for this analysis. 
For clarity, one of the paddle counters and the two outer ring counters on each
side were omitted and the positions of ring 
and paddle counters are not to scale. }
\label{1}
\end{figure}
Details of the trigger performance and the event selection procedure can 
be found in \cite{phobosprl,phoboscent}. 
Based on MC simulations, we estimate that the trigger was sensitive to  
$97\pm 3$\% of the total hadronic inelastic cross section. For the selection of 
central events used here, no contamination from beam-gas 
or other background collisions was found.
The distribution of the combined signal from PP and PN (``paddle mean'') 
is shown in Fig.~2 for events with more than two hits in each paddle counter.
This selection includes $88 \pm 3$\% of the hadronic cross section. 
We selected the top 6\% of the total hadronic cross 
section with the largest paddle mean, corresponding to the most central collisions 
with the largest number of participating nucleons.
Applying the 6\% centrality cut to MC events, we deduce that the average 
number of participating nucleons in the data is 
$\langle N_{part} \rangle = 344 \pm 10$~(syst).
The systematic uncertainty was estimated using MC studies \cite{phoboscent}
and an analysis of the correlations between the signals in PN and PP.

For the selected central events, the event vertex was determined using the 
SPEC and VTX subdetectors. Details of the vertex finding procedure can be 
found in \cite{phobosprl,phoboscent}. 
MC studies and correlations between the vertex positions
found independently using SPEC and VTX detectors indicate a vertex resolution
of better than $400\mu$m and a vertex finding efficiency of 100\% in a fiducial range of 
$-10~\mbox{cm} < z_{vtx} < 10 \mbox{ cm}$. 
A total of 639 central events were selected in this vertex range.

\begin{figure}[t]
\centerline{
\epsfig{file=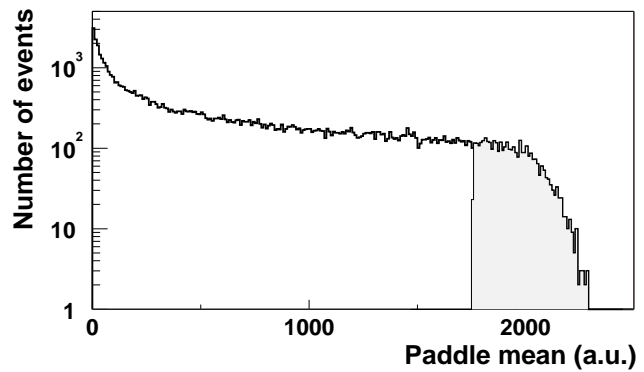,width=9cm}
}
\caption{Paddle signal distribution for Au+Au events at 
$\sqrt{s_{_{NN}}} = $ 200~GeV. The shaded area 
indicates the region selected by the 6\% centrality cut.}
\label{Figure3}
\end{figure}
The charged particle pseudorapidity density was determined using four
independent methods, which have been discussed in detail in previous 
publications \cite{phobosprl,phoboscent,phobosdndeta}. 
We used an analysis of two-hit combinations (tracklets)
in SPEC and VTX 
detectors \cite{phobosprl,phoboscent}, a hit-counting method  
in OCT and RING \cite{phobosdndeta} and a method based on the 
observed energy loss 
in the silicon sensors (``analog method'') \cite{phobosdndeta} 
for all subdetectors. 
For all four methods the multiplicity densities 
$dN_{ch}/d\eta |_{|\eta|<1}$ were corrected for particles
which stop in the material of the beam pipe and the first
detector layer, particles from secondary interactions and feed-down 
products from weak decays of neutral strange particles. 
The resulting systematic errors on $dN_{ch}/d\eta |_{|\eta|<1}$ for 
each method at $\sqrt{s_{_{NN}}} = $ 130~GeV are described in 
\cite{phobosprl,phoboscent,phobosdndeta}
and range from 4.5\% for the SPEC tracklet analysis to 10\% for the 
hit-counting and analog methods. MC studies showed that the systematic uncertainties
are essentially the same at $\sqrt{s_{_{NN}}} =$ 200~GeV.
The combined result of all four methods is obtained using
the inverse square of the estimated uncertainties as weights in the 
average.  Based on MC studies and the comparison of all four 
independent multiplicity analyses, we estimate the 
overall systematic uncertainty of the combined result to be less than 6\%. 
The statistical error is negligible.
We obtain a primary charged particle density of 
$dN_{ch}/d\eta |_{|\eta|<1}  = 650 \pm 35 \mbox{(syst)}$
for the 6\% most central Au+Au collisions at $\sqrt{s_{_{NN}}} = $ 200~GeV. 
Normalizing per participant pair, we find
$dN_{ch}/d\eta |_{|\eta|<1} / \langle \frac{1}{2} N_{part} \rangle = 
3.78 \pm 0.25 \mbox{(syst)}$.

\begin{figure}[t]
\centerline{ \epsfig{file=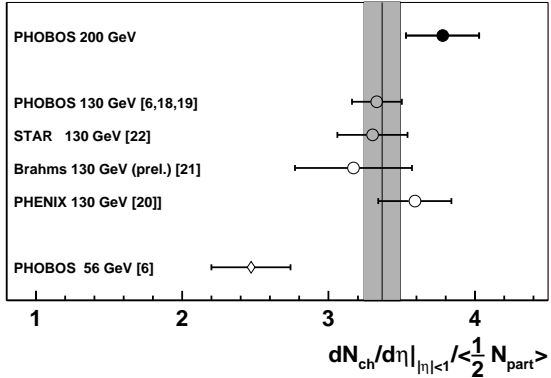,width=9cm} }
\caption{Summary of RHIC results on the pseudorapidity density normalized per 
participant pair for central Au+Au collisions at 
$\sqrt{s_{_{NN}}} = $ 
56, 130 and 200~GeV. The line and grey area show the averaged result 
from all 4 experiments at $\sqrt{s_{_{NN}}} = $ 130 GeV.}
\label{Figure3b}
\end{figure}

A compilation of all results for 
$dN_{ch}/d\eta |_{|\eta|<1}/\langle \frac{1}{2} N_{part}\rangle$ 
in  central Au+Au collisions obtained at RHIC is shown in Fig~3. It includes results 
from PHOBOS at $\sqrt{s_{_{NN}}} =$ 56~GeV \cite{phobosprl}, 
from  all four RHIC experiments at $\sqrt{s_{_{NN}}} =$ 130~GeV 
\cite{phobosprl,phoboscent,phobosdndeta,phenix_cent,brahms_qm2001,star_mult} 
and our new result at $\sqrt{s_{_{NN}}} =$ 200~GeV. The average value 
at 130~GeV is $dN_{ch}/d\eta |_{|\eta|<1}/\langle \frac{1}{2} N_{part} \rangle = 3.37 
\pm 0.12$, compared to $3.78 \pm 0.25$ reported here for 200~GeV.

\begin{figure}[t]
\centerline{ \epsfig{file=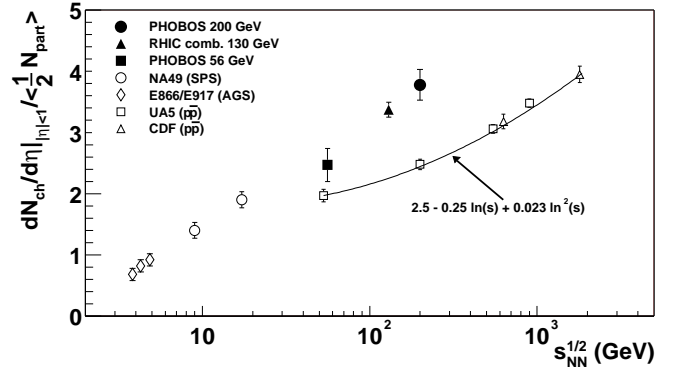,width=9cm} }
\caption{Energy dependence of the pseudorapidity density normalized per
participant pair for central nucleus-nucleus collisions.
The data are compared with $p\overline{p}$ data and nucleus-nucleus data 
from lower energies}
\label{Figure4}
\end{figure}

In Fig.~\ref{Figure4} the normalized yield per participant is compared
to central Au+Au and Pb+Pb collisions at lower energies, 
including fixed target experiments \cite{e917,na49,na49_qm01}.
Also shown for comparison 
are results from proton-antiproton ($p\overline{p}$) collisions \cite{pp} and 
an interpolation of the $p\overline{p}$ data.
The $dN_{ch}/d\eta$ values from
E866/E917 \cite{e917} and  
NA49 \cite{na49,na49_qm01} 
were obtained by integrating the measured 
charged particle rapidity and $p_T$-distributions.
Within the precision of the existing data, an approximately logarithmic rise
of $dN_{ch}/d\eta |_{|\eta|<1}/\langle \frac{1}{2} N_{part} \rangle$ with $\sqrt{s_{_{NN}}}$ 
is observed over the full range of collision energies.

A direct measurement of the ratio $R_{200/130}$ of normalized 
multiplicity densities $dN_{ch}/d\eta |_{|\eta|<1}/\langle \frac{1}{2} N_{part}\rangle$ 
at $\sqrt{s_{_{NN}}} =$ 200~GeV and 130~GeV was also performed. 
\begin{figure}[t]
\centerline{ \epsfig{file=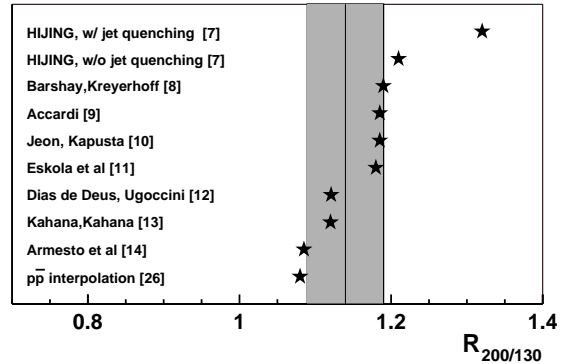,width=9cm} }
\caption{Ratio $R_{200/130}$ of pseudorapidity densities 
$dN_{ch}/d\eta |_{|\eta|<1}$ at  $\sqrt{s_{_{NN}}} =$ 200~GeV and 130 GeV. 
The predicted increase from 
various models is compared to the observed increase in the data 
(vertical line). The grey band shows the 90\% confidence interval.} 
\label{5}
\end{figure}
\noindent
The ratio was determined separately for each of the four multiplicity 
analyses at 130 and 200 GeV and then averaged, weighted with the 
inverse square of the systematic error for each ratio measurement. Many of 
the systematic uncertainties in each of the analyses partially cancel,
leading to a significant reduction in the uncertainty of the ratio.
The theoretical uncertainty in the $N_{part}$ calculation is 
also largely eliminated in the ratio. We estimate that the uncertainty 
from the energy dependence of absorption, feed-down and acceptance corrections
is less than 0.03 in $R_{200/130}$. Based on MC tests and the comparison
of the results for the four different methods 
we estimate a total systematic uncertainty of less than 0.05 for the 
averaged result.
The combined estimate of the ratio 
of normalized multiplicity densities 
$dN_{ch}/d\eta |_{|\eta|<1}/\langle \frac{1}{2} N_{part}\rangle$ at 
$\sqrt{s_{_{NN}}} =$ 200~GeV and 130~GeV, obtained from all four 
multiplicity analysis methods, is $R_{200/130} = 1.14 \pm 0.05$.
The quoted uncertainty is entirely systematic and corresponds to a 
90\% confidence level.
In Fig.~5 this value is compared with the $p\overline{p}$ 
parametrization \cite{pp} and several model calculations. 
The expected increase from the interpolation of 
$p\overline{p}$ results  falls slightly below the allowed region.
Within the systematic uncertainty, the result is in
agreement with most other model predictions. 
However, we do not observe the large increase 
expected in the HIJING model \cite{wang_0008014}, 
which predicted $R_{200/130} = 1.32$ using the default jet quenching parameters.
Further theoretical work 
will be required to describe both the preliminary results on 
large $p_T$ spectra, as well as the energy dependence of particle multiplicities.
\begin{figure}[t]
\centerline{ \epsfig{file=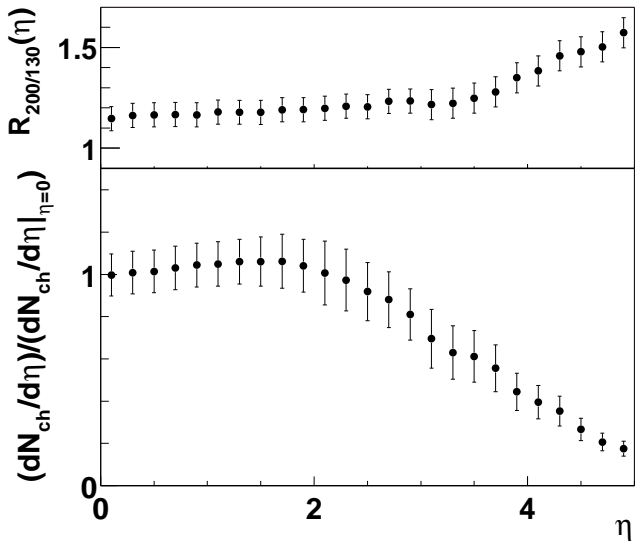,width=9cm} }
\caption{The lower plot shows the shape of $dN_{ch}/d\eta$ at $\sqrt{s_{_{NN}}} =$ 200~GeV, obtained from 
the analog method, normalized to $dN_{ch}/d\eta|_{\eta=0}$. 
On top we show the pseudorapidity 
dependence of $R_{200/130}$ using the result of the analog  
method at both energies. In both plots, the error bars indicate the systematic 
uncertainty. }
\label{6}
\end{figure}

Additional information about the change in particle production from
130 GeV to 200 GeV collision energy can be gained from Fig.~6. Using the
results from the analog method \cite{phobosdndeta}, it  
shows the scaled $dN_{ch}/d\eta$ out to $\eta=5$ for central collisions at 200 GeV (bottom)
and the ratio of the pseudorapidity distributions at 200 and 130~GeV (top).
A slow increase of the ratio is observed to $\eta\approx 3.5$. This is 
followed by a steeper rise at higher $\eta$, corresponding to a widening of the 
pseudorapidity distribution and reflecting
the shift in the fragmentation region due to the larger beam rapidities
at 200~GeV.

In summary, we have performed the first measurement of the 
pseudorapidity density of primary charged particles in 
Au+Au collisions at $\sqrt{s_{_{NN}}} =$ 200~GeV.
Near mid-rapidity, this density shows an approximately logarithmic
evolution over a broad range of collision energies.
Going from $\sqrt{s_{_{NN}}} =$ 130~GeV to 200~GeV, the observed 
increase is a factor of $1.14 \pm 0.05$, 
corresponding to a moderate increase in initial energy density.
Over this energy interval, a smooth evolution of $dN_{ch}/d\eta$ in all 
regions of pseudorapidity is
seen. These results give further constraints for models including 
effects of the partonic medium in the early collision stages.

Acknowledgements: 
This work was partially supported by US DoE grants DE-AC02-98CH10886,
DE-FG02-93ER40802, DE-FC02-94ER40818, DE-FG02-94ER40865, DE-FG02-99ER41099, W-31-109-ENG-38.
NSF grants 9603486, 9722606 and 0072204. The Polish groups were partially supported by KBN grant 2 P03B
04916. The NCU group was partially supported by NSC of Taiwan under 
contract NSC 89-2112-M-008-024.

\end{document}